\documentclass[a4paper,11pt]{article}
\usepackage{pos}
\usepackage{pgfplots}
\usepackage{relsize}
\usepackage{slashed}
\usepackage{wrapfig}
\usepackage{graphicx} 
\usepackage{subcaption}

\renewcommand{\logo}{\relax}

\title{Searches for new physics models via the same-sign diboson (SSdB) + 
 $E_T \!\!\!\!\!\!/$~~ and precise measurement of top quark features at the LHC}
\author*[a]{Dibyashree Sengupta}
\affiliation[a]{
INFN, Laboratori Nazionali di Frascati, 
\\ 
Via E. Fermi 54, 00044 Frascati (RM), Italy}

\emailAdd{Dibyashree.Sengupta@lnf.infn.it}
\abstract{Till today, although the Standard Model (SM) is the most celebrated theory that explains nature almost completely, there are still some phenomena observed in nature that the SM cannot explain. That is why it is needed to look for theories beyond the Standard Model (BSM). While the ATLAS/CMS experiments discovered a Standard Model-like Higgs boson at the Large Hadron Collider (LHC), no compelling new physics signal has been seen yet. Several searches have been performed at the LHC to look for new physics signal. One such novel signal is the same-sign diboson (SSdB) + $E_T \!\!\!\!\!\!/$~~ which is a rather clean signal with negligibly small SM background. Such a unique signature can be observed in more than one well-motivated BSM scenarios, namely: (i) natural SUSY models, (ii) type-III seesaw model and (iii) type-II seesaw/Georgi-Machacek model. In the first part of this poster I present the discovery prospects of this signal that has been analyzed in these BSM models in current and future runs of the LHC beside providing ways to distinguish among these different BSM models. Furthermore, the LHC, being a ``top quark factory", helps in precise measurement of various properties of the top quark. Deviation from the SM prediction in measuring these properties of the top quark can, very efficiently, shed light on new physics signal. In the second part of this poster I present a work in progress where we aim to show how precise measurement of quantities related to top quark features can indicate towards a new physics signal.}

\FullConference{The European Physical Society Conference on High Energy Physics (EPS-HEP2023)\\
 21-25 August 2023\\
Hamburg, Germany\\}

\begin{document}
\maketitle

\section{Introduction}
With the discovery of the Higgs Boson~\cite{Aad:2012tfa, Chatrchyan:2012ufa} the Standard Model (SM) reached its pinnacle of success. However, it cannot explain several phenomena observed in nature. Hence Beyond Standard Model (BSM) scenarios are needed. There have been many experimental searches for several BSM signals with no significant success yet. Here we study a novel new physics signal, namely, the same-sign diboson (SSdB) + {$\slashed{E}_{T}$} signature which can be observed in more than one well-motivated BSM scenarios, namely: (i) natural supersymmetric (SUSY) models~\cite{Matalliotakis:1994ft, Baer:2005bu, Baer:2016hfa, Randall:1998uk, Baer:2018hwa, Baer:2020kwz, Baer:2016lpj}, (ii) type-III seesaw model~\cite{Foot:1988aq}, and (iii) type-II seesaw~\cite{Magg:1980ut, Schechter:1980gr, Mohapatra:1979ia, Lazarides:1980nt}/Georgi-Machacek model~\cite{Georgi:1985nv}. The discovery prospects of this signal has been analyzed in these BSM models in the Large Hadron Collider (LHC) beside distinguishing among these different BSM models in Sec.~\ref{sec:models}. The LHC being a "top quark factory", aids in a careful study of top quark properties. So we use it to study a particular observable, namely, the invariant mass of the b-jet and the lepton ($m_{b\ell}$) and check if the presence of any BSM particle with mass close to that of top quark can cause any change in the SM prediction of $m_{b\ell}$. In Sec.~\ref{sec:top}, an ongoing work is presented where we show that considering MSSM as a signal, after a set of acceptance cuts, namely cut set A, we do get some excess in signal in the low $m_{b\ell}$ region in the $m_{b\ell}$ distribution~\cite{top:2023bcfs}. Finally we conclude in Sec.~\ref{sec:conclusion}.

%Furthermore, the LHC, being a ``top quark factory", helps in precise measurement of various properties of the top quark. Deviation from the SM prediction in measuring these properties of the top quark can, very efficiently, shed light on new physics signal. In Sec.~\ref{sec:top} I  present a work in progress where we aim to show how precise measurement of quantities related to top quark features can indicate towards a new physics signal. 

\section{BSM scenarios yielding same-sign diboson (SSdB) + {$\slashed{E}_{T}$} signature}
\label{sec:models}
 The charged bosons in the final state decay leptonically. Hence the final state is same-sign dilepton (SSdL)+ $\slashed{E}_{T}$. For  simulations, {\tt MadGraph5$\_$aMC@NLO}~\cite{Alwall:2011uj,Alwall:2014hca} interfaced with {\tt Pythia}~8.2 ~\cite{Sjostrand:2014zea} and {\tt Delphes}~3.4.2~\cite{deFavereau:2013fsa} with the anti-$k_T$ jet algorithm~\cite{Cacciari:2008gp} has been used. We generate the Les Houches Accord (LHA) file for the NUHM2 signal using {\tt Isajet}~7.88~\cite{Paige:2003mg}. The K-factors for the signal and SM background processes are considered as in Ref.~\cite{Chiang:2021lsx, Baer:2017gzf}. 

\subsection{Natural SUSY models}

\begin{figure} [h!]
\begin{center}
\includegraphics [width=5cm,height=3cm] {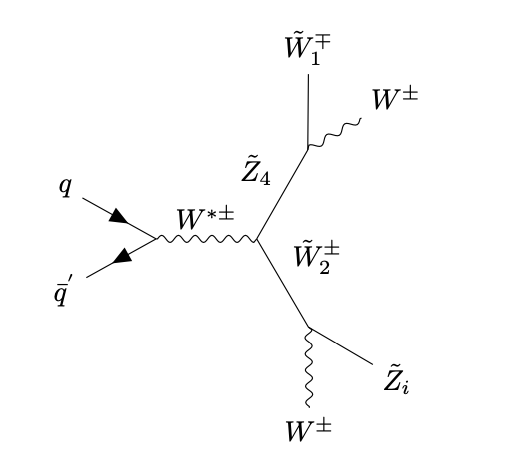}
\end{center}
\vspace*{-0.3in}
\caption{Feynman diagram for SSdB + {$\slashed{E}_{T}$} production at the LHC in Natural SUSY models}
\label{fig:susy}
\end{figure}

Supersymmetry (SUSY) ~\cite{Baer:2006rs} solves a number of questions that the SM cannot making the former a well-motivated BSM theory. However, current LHC data ~\cite{Aaboud:2017vwy, Vami:2019slp,Vami:2019slp, ATLAS:2019oho, CMS:2019ysk} has pushed the lower bounds on the masses of sparticles in the multi-TeV regime thereby questioning the naturalness of weak scale SUSY~\cite{Craig:2013cxa} in light of the older notions of naturalness~\cite{Barbieri:1987fn, Papucci:2011wy, Kitano:2006gv}. However, these earlier notions of naturalness can be updated to a more conservative electroweak naturalness measure~\cite{Baer:2013gva, Mustafayev:2014lqa, Baer:2014ica, Baer:2012cf} which allow a few Natural SUSY models~\cite{Matalliotakis:1994ft, Baer:2005bu, Baer:2016hfa, Randall:1998uk, Baer:2018hwa, Baer:2020kwz, Baer:2016lpj} whose parameter space is yet to be completely probed by the LHC~\cite{Baer:2018hpb}. All of these Natural SUSY models are characterized by a higgsino-like Lightest supersymmetric particle (LSP) which, considering R-parity conservation, serves as a good cold dark matter (CDM) and hence appear as {$\slashed{E}_{T}$} in the LHC. One such Natural SUSY model, namely, the the two extra parameter non-universal Higgs (NUHM2) model~\cite{Matalliotakis:1994ft, Baer:2005bu} has been considered here that gives rise to the SSdB + {$\slashed{E}_{T}$} signature via wino pair production as shown in Fig.~\ref{fig:susy}. 

After the A3 (A3$^{\prime}$)-cuts: SSdL, no b-jets, $p_T(\ell_1) > 20$~GeV, $\slashed{E}_{T} > 250 (350)$~GeV,  $m_{T_{\rm min}} > 200 (325)$~GeV at $\sqrt{s} = 27 (100)$ TeV, the NUHM2 model stands out as the chosen benchmark points yield the following significance: 
NUHM2 : 8.06 (13.6) at $\mathcal{L} = 3$ ab$^{-1}$ and 18.01(30.5) at $\mathcal{L} = 15$ ab$^{-1}$;
Type III : 1.21 (1.5) at $\mathcal{L} = 3$ ab$^{-1}$ and 2.71(3.3) at $\mathcal{L} = 15$ ab$^{-1}$;
GM : 0.0135 (0.06) at $\mathcal{L} = 3$ ab$^{-1}$ and 0.03(0.14) at $\mathcal{L} = 15$ ab$^{-1}$.

\subsection{Type-III seesaw model}
\begin{figure} [h!]
\begin{center}
\includegraphics [width=5cm,height=3cm] {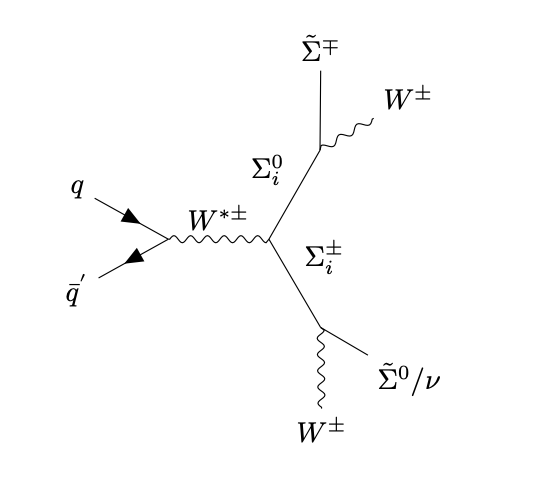}
\end{center}
\vspace*{-0.3in}
\caption{SSdB + $\slashed{E}_{T}$ signature at the LHC in the type-III seesaw model}
\label{fig:typeiii}
\end{figure}

The type-III seesaw model~\cite{Foot:1988aq} is a well-motivated BSM theory as it explains neutrino masses and mixings by extending the SM spectrum by three generations of $SU(2)_L$ triplet fermions with hypercharge $Y=0$, the lightest of which (denoted by  $\tilde{\Sigma}$) with mass around a few hundred~GeV can have a lifetime long enough to escape detection~\cite{Jana:2019tdm} and hence shows up as large $\slashed{E}_{T}$  in the LHC. We consider the heavier two generations to be mass degenerate for simplicity. Thus the type-III seesaw model yields the the SSdB + $\slashed{E}_{T}$ signature via Fig.~\ref{fig:typeiii}. 

After the B2 (B2$^{\prime}$)-cuts: SSdL, no b-jets, $p_T(\ell_1) > 20$~GeV, njet $\le 1$ + $\slashed{E}_{T} > 100 (120)$~GeV + 105~GeV~$< m_{T_{\rm min}} < 195$~GeV + $200$~GeV $<$ MCT $<$ $325 (350)$~GeV at $\sqrt{s} = 27 (100)$ TeV, the type-III seesaw model model stands out as the chosen benchmark points yield the following significance: 
NUHM2 : 0.52 (0.8) at $\mathcal{L} = 3$ ab$^{-1}$ and 1.2(1.8) at $\mathcal{L} = 15$ ab$^{-1}$;
Type III : 3.5 (4.3) at $\mathcal{L} = 3$ ab$^{-1}$ and 7.8(9.6) at $\mathcal{L} = 15$ ab$^{-1}$ ;
GM : 0.45 (1.4) at $\mathcal{L} = 3$ ab$^{-1}$ and 1.0(3.1) at $\mathcal{L} = 15$ ab$^{-1}$.

\subsection{Type-II seesaw/Georgi-Machacek model}
\begin{figure}
\begin{subfigure}[h]{0.4\linewidth}
\includegraphics[width=5cm,height=3cm]{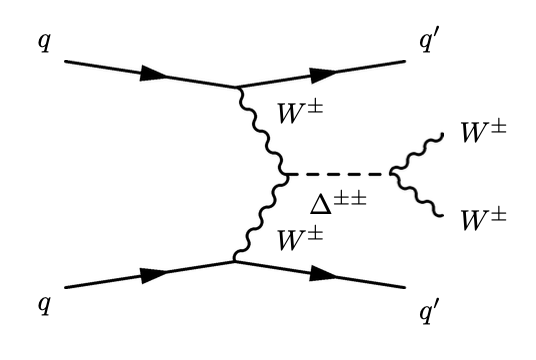}
\caption{}
\label{fig:feyn4}
\end{subfigure}
\hfill
\begin{subfigure}[h]{0.4\linewidth}
\includegraphics[width=0.7\linewidth]{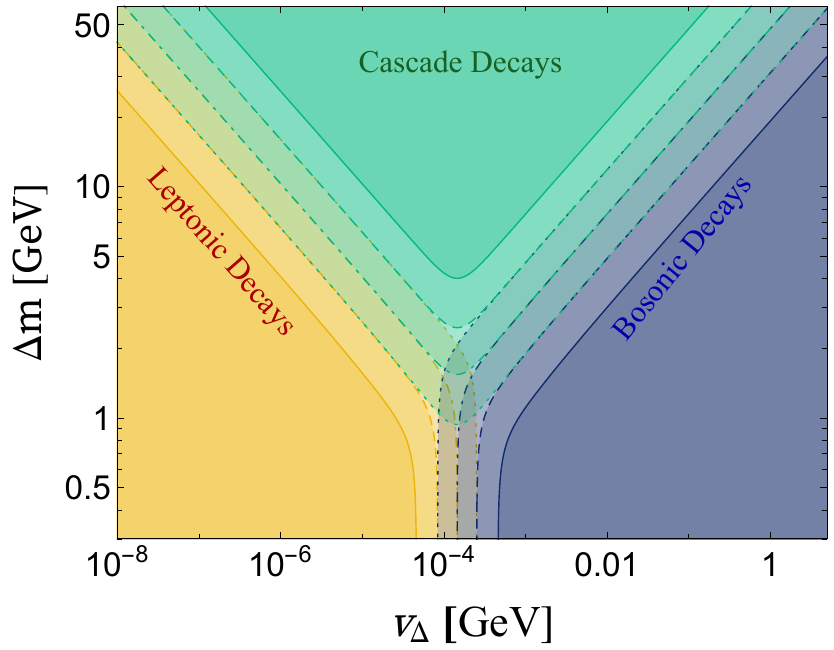}
\caption{}
\label{fig:phase}
\end{subfigure}%
\caption{(a)SSdB + forward jets production at LHC in the type-II seesaw/GM models.(b)Decay phase diagram of doubly-charged scalar ($\Delta^{\pm \pm}$) with mass = 300~GeV.}
\label{fig:typeii}
\end{figure}

A pair of same-sign bosons can originate from the decay of a doubly-charged scalar ($\Delta^{++}$) which is present in many BSM scenarios~\cite{Magg:1980ut, Schechter:1980gr, Lazarides:1980nt, Mohapatra:1980yp, Pati:1974yy, Mohapatra:1974hk, Senjanovic:1975rk, Kuchimanchi:1993jg, Babu:2008ep, Babu:2014vba, Basso:2015pka, Zee:1985id, Babu:1988ki, ArkaniHamed:2002qx, Babu:2020hun, Georgi:1985nv, Gunion:1989ci, Babu:2009aq, Bonnet:2009ej, Bhattacharya:2016qsg, Kumericki:2012bh} one of which is the simplest type-II seesaw model~\cite{Magg:1980ut, Schechter:1980gr, Mohapatra:1979ia, Lazarides:1980nt}. Since $\Delta^{++}$ is produced through VBF, as in Fig.~\ref{fig:typeii}\subref{fig:feyn4}, the production cross-section is extremely small in the type-II seesaw model due to the $T$-parameter constraint~\cite{Magg:1980ut, Schechter:1980gr, Mohapatra:1979ia, Lazarides:1980nt, Zyla:2020zbs} which can be enhanced in the Georgi-Machacek(GM) model~\cite{Georgi:1985nv}, thanks to the custodial symmetry~\cite{Chiang:2012dk, Chiang:2015amq, Blasi:2017xmc, Chiang:2018xpl}. The parameters of $\Delta^{++}$ is chosen such that it falls in the blue region in Fig.~\ref{fig:typeii}\subref{fig:phase} and hence primarily decays into $W^{+}$ bosons~\cite{Melfo:2011nx,Aoki:2011pz,Chiang:2012cn}. Considering leptonic decay of the same-sign dibosons and the jets being forward, the final state mimics our signature of interest.

After the C3 (C3$^{\prime}$)-cuts: SSdL, no b-jets, $p_T(\ell_1) > 20$~GeV, MCT $\le 300$~GeV + njet $\ge$ 2, $\Delta \eta (j_1, j_2) > 5$, $\slashed{E}_{T}$ $>$ 50~GeV, $m_{T_{\rm min}}$ $>$ 105 (120) ~GeV at $\sqrt{s} = 27 (100)$ TeV, the GM model stands out as the chosen benchmark points yield the following significance: 
NUHM2 : 0 (0.22) at $\mathcal{L} = 3$ ab$^{-1}$ and 0(0.48) at $\mathcal{L} = 15$ ab$^{-1}$;
Type III : 0.22 (1.23) at $\mathcal{L} = 3$ ab$^{-1}$ and 0.5(2.7) at $\mathcal{L} = 15$ ab$^{-1}$ ;
GM : 2.5 (3.02) at $\mathcal{L} = 3$ ab$^{-1}$ and 5.5(6.75) at $\mathcal{L} = 15$ ab$^{-1}$.

\section{New physics from measurement of top quark features}
\label{sec:top}
Here we consider top quark pair production and study a particular observable, namely, the invariant mass of the b-jet and the lepton ($m_{b\ell}$) which has been used to extract the mass of the top quark~\cite{CMS:2014cza, CMS:2009kup}. Our goal is to check if there exist a BSM particle with mass around $m_t$, can it cause any deviation in the $m_{b\ell}$ distribution~\cite{Franceschini:2017eyj}. Although this method applies to any BSM theory that can accommodate a new particle with mass around $m_t$, we consider the MSSM as an example with the mass of the lightest stop ($\tilde{t_1}$) to be around $m_t$. $\tilde{t_1}$ is pair-produced and decayed as: $p p > \tilde{t_1} \tilde{t_1}^*, (\tilde{t_1} > b \tilde{\chi_1^{+}}, \tilde{\chi_1^{+}} > \tilde{\chi_1^{0}} W^+/W^{*+} > \tilde{\chi_1^{0}} \ell^+ \nu_l), (\tilde{t_1}^* > \bar{b} \tilde{\chi_1^{-}},  \tilde{\chi_1^{-}} >  \tilde{\chi_1^{0}} W^-/W^{*-} > \tilde{\chi_1^{0}} \ell^- \bar{\nu_l})$ where $\tilde{\chi_1^{\pm}}$ is the lightest chargino and $\tilde{\chi_1^{0}}$ is the lightest neutralino i.e., the LSP. Here we consider $\tilde{t_1}$ decaying to $ b \tilde{\chi_1^{+}}$ with 100$\%$ branching ratio and do not consider the decay channel $\tilde{t_1} > t \tilde{\chi_1^{0}}$ because the latter has been already excluded by experiments~\cite{CMS:2019wav}. Therefore, the signal yields the same final state as top-quark pair production and their $m_{b\ell}$ distributions are compared. Three cases with $m_{\tilde{t_1}}$ = $180$, $200$ and $220$~GeV have been investigated~\cite{top:2023bcfs}. For simplicity, here we present the results for $m_{\tilde{t_1}}$ = $200$~GeV only.

%\begin{figure} [h]
%\begin{center}
%\includegraphics [width=0.2\textwidth] {top_italy_feyn.png}
%\end{center}
%\caption{MSSM signal where each top squark ($\tilde{t_1}$) decay to $b$ and the lightest chargino $\tilde{\chi_1^{\pm}}$ which again decays leptonically to the LSP $\tilde{\chi_1^{0}}$}
%\label{fig:mssm}
%\end{figure}    

\subsection{Allowed parameter space points}
\begin{figure} [h!]
\begin{center}
\includegraphics [width=5cm,height=3cm] {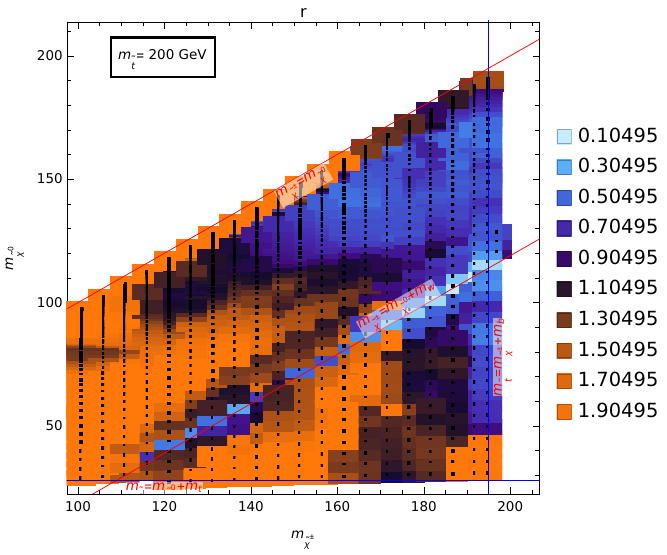}
\end{center}
\vspace*{-0.3in}
\caption{$r<1$ implies that a parameter space point is allowed by the current experimental constraints}
\label{fig:r}
\end{figure}

 We have used {\tt SPheno-}~4.0.3~\cite{Porod:2003um, Porod:2011nf} interfaced with {\tt SARAH-}~4.15.1~\cite{Staub:2008uz, Staub:2013tta, Staub:2012pb, Staub:2010jh, Staub:2009bi, Staub:2011dp, Dreiner:2012dh, Porod:2014xia, Goodsell:2014bna, Staub:2017jnp, Goodsell:2017pdq, Goodsell:2018tti, Staub:2015kfa, Goodsell:2015ira, Goodsell:2016udb} to generate several such parameter space points with $m_{\tilde{t_1}}$ $\sim$ 200 GeV and different values of $m_{\tilde{\chi_1^{\pm}}}$ and $m_{\tilde{\chi_1^{0}}}$. We passed each of these points through {\tt smodels-}~2.2.1~\cite{Alguero:2021dig} which returns a value of $r$ by comparing the point with all existing experimental searches. A parameter space point is said to be allowed by the experimental constraints if $r<1$. In Fig.~\ref{fig:r}, we show the $r$ values for each parameter space point for $m_{\tilde{t_1}}$ $\sim 200$~GeV. Finally each of these points have been passed through {\tt Pythia}~8.3 ~\cite{Bierlich:2022pfr} to evaluate the cross-section and significance as discussed in Sec.~\ref{sec:sig}

\subsection{Significance}
\label{sec:sig}
\begin{figure} [h!]
\begin{subfigure}[h]{0.3\linewidth}
\includegraphics[width=\linewidth]{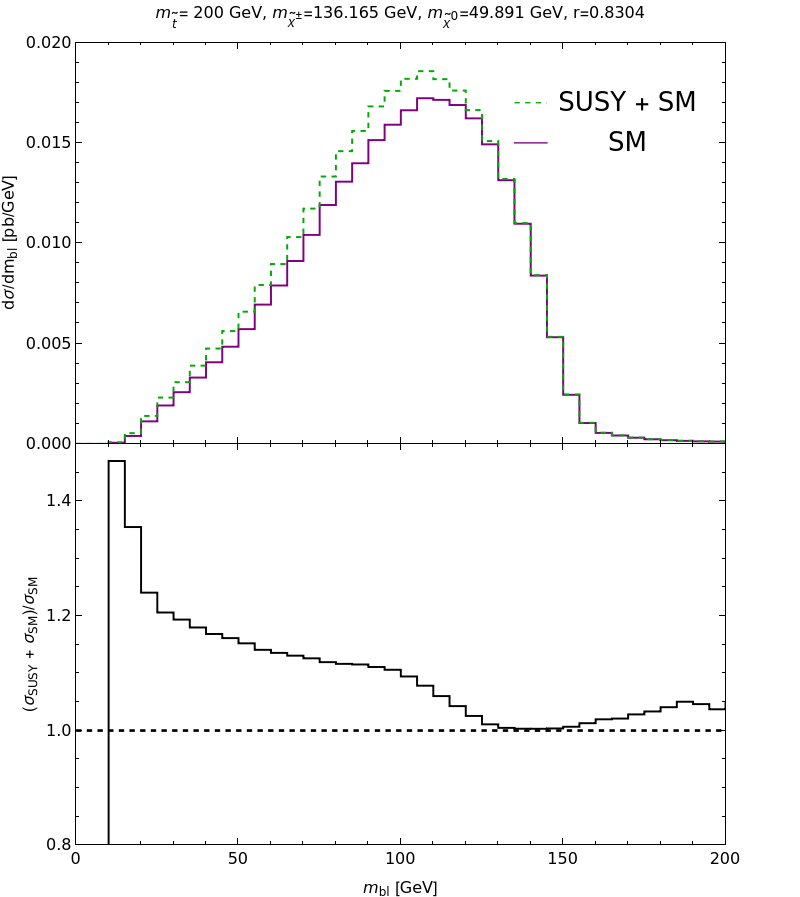}
\caption{$m_{b\ell}$ distribution at $\sqrt{s} = 13$~TeV after cut set A}
\label{mbl_bm2}
\end{subfigure}
\hfill
\begin{subfigure}[h]{0.3\linewidth}
\includegraphics[width=\linewidth]{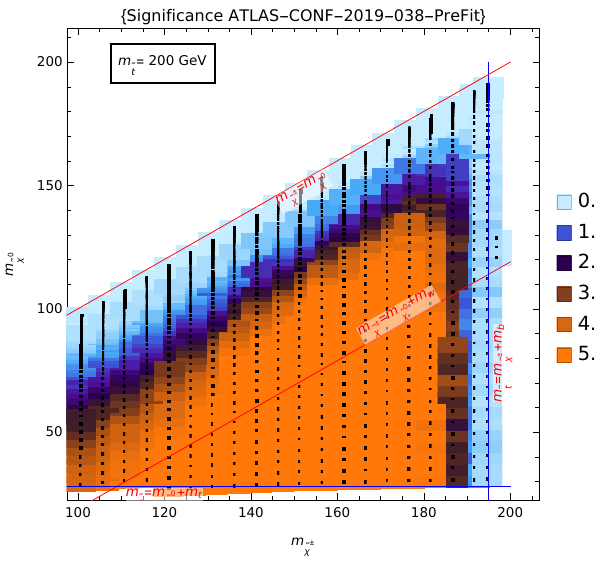}
\caption{Significance = $S/(B*u_{B}^{ATLAS})$}
\label{atlasprefit}
\end{subfigure}
\hfill
\begin{subfigure}[h]{0.3\linewidth}
\includegraphics[width=\linewidth]{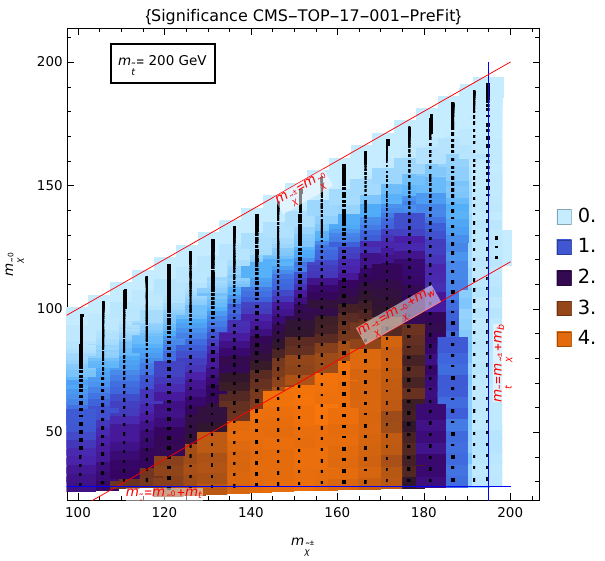}
\caption{Significance = $S/(B*u_{B}^{CMS})$}
\label{cmsprefit}
\end{subfigure}%
\caption{$m_{\tilde{t_1}}$ $\sim 200$~GeV}
\label{fig:mstop200}
\end{figure}

Fig.~\ref{fig:mstop200}\subref{mbl_bm2} shows the $m_{b\ell}$ distribution after cut set A: $p_T(\ell) > 25$~GeV, $|\eta(\ell)| < 2.5$, $R(j) = 0.4$, $p_T(j) > 25$~GeV, $|\eta(j)| < 2.5$, $\Delta_R(\ell j)>0.2$~\cite{ATLAS:2019onj, ATLAS:2017vgz, CMS:2016hdd} at $\sqrt{s} = 13$~TeV  for a  benchmark point with $m_{\tilde{t_1}}$ $\sim 200$~GeV,  $m_{\tilde{\chi_1^{\pm}}}$ $\sim 136.2$~GeV and $m_{\tilde{\chi_1^{0}}}$ $\sim 49.9$~GeV with $r<1$ and reflects some signal excess in the low $m_{b\ell}$ region. This benchmark point has significance $\sim 10$ and $\sim 4$ calculated by summing in quadrature over all bins of the $m_{b\ell}$ distribution the quantity $S/(B*u_B)$, when the relative uncertainity in the background, denoted by $u_B$, is extracted from ATLAS~\cite{ATLAS:2019onj} and CMS~\cite{CMS:2018fks} respectively, where $S=$ number of signal events and $B=$ number of background events in a particular bin in the $m_{b\ell}$ distribution as obtained from the simulations in {\tt Pythia}~8.3 ~\cite{Bierlich:2022pfr}. In Fig.~\ref{fig:mstop200}\subref{atlasprefit} and Fig.~\ref{fig:mstop200}\subref{cmsprefit} we show the significance, evaluated as discussed above,  of various parameter space points varying in $m_{\tilde{\chi_1^{\pm}}}$ and $m_{\tilde{\chi_1^{0}}}$ but having $m_{\tilde{t_1}}$ $\sim 200$~GeV with the relative uncertainity in background, $u_B$, extracted from ATLAS~\cite{ATLAS:2019onj} and CMS~\cite{CMS:2018fks} respectively. 

\section{Conclusion}
\label{sec:conclusion}
We find that it is possible to observe the SSdB + ${E}_{T}{\!\!\!\!/}$ signature in three well-motivated BSM scenarios, namely: (i) NUHM2 model (ii) type-III seesaw model and (iii) type-II seesaw/Georgi-Machacek model; the first being visible at IL = 3ab$^{-1}$ while the latter two at IL = 15ab$^{-1}$. We chose a simple MSSM signal with $m_{\tilde{t_1}}$ $\sim m_{t}$ such that the $\tilde{t_1}$ mimics the $t$ decay assuming leptonic decay of $W^{\pm}$ boson so as to compare their $m_{b\ell}$ distribution. At $\sqrt{s} = 13$~TeV after the cut set A some excess in the signal in the $m_{b\ell}$ distribution has been observed for an allowed benchmark point. We also find a number of allowed parameter space points that have high significance.

\acknowledgments{I thank Universität Hamburg and the organizers of The European Physical Society Conference on High Energy Physics (EPS-HEP) 2023 for their kind hospitality. I thank my collaborators Cheng-Wei Chiang, Sudip Jana, Gennaro Corcella, Roberto Franceschini and Emanuele Angelo Bagnaschi. I thank Howard Baer for useful discussions.  The first part of this research was supported by the Ministry of Science and Technology (MOST) of Taiwan under Grant Nos.~108-2112-M-002-005-MY3 and 109-2811-M-002-570.}

\bibliographystyle{JHEP}

\bibliography{reference}

\end{document}